\documentclass[%
 aip,
  jor,
 amsmath,amssymb,
 reprint,%
]{revtex4-1}
\usepackage{graphicx}  % needed for figures
\usepackage{dcolumn}   % needed for some tables
\usepackage{bm}        % for math
\usepackage{amssymb}   %for math
\usepackage{subfigure}
\usepackage{float}
\usepackage{blindtext}
\usepackage{color}
\usepackage{xr}
\usepackage{newfloat}
\usepackage{wasysym}
\usepackage{MnSymbol}
\usepackage{multirow,bigdelim}
\usepackage{rotating}

\usepackage[utf8]{inputenc}
\usepackage[T1]{fontenc}
\usepackage{color}
\definecolor{red}{rgb}{1,0,0}
\definecolor{blue}{rgb}{0,0,1}
\definecolor{black}{rgb}{0,0,0}

\begin{document}

\preprint{}

\title{Oscillatory shear flows of dense suspensions at imposed pressure: Rheology and microstructure} 
\author{Junhao Dong}
\author{Martin Trulsson}
\email{martin.trulsson@teokem.lu.se}
\affiliation{Theoretical Chemistry, Department of Chemistry, Lund University, Sweden}
\date{\today}

\begin{abstract}

Oscillatory shear has been widely used to study the rheological properties of suspensions under unsteady shear. Furthermore, recent works have shown that oscillatory flows can improve the flowability of dense suspensions. While most studies have been done under constant volume, we here study oscillatory shear flows of two-dimensional suspensions using a normal pressure-controlled set-up.
To characterise the rheology, we introduce both a complex macroscopic friction coefficient $\mu^{*}$, following the convention of the complex viscosity $\eta^{*}$, and a shear-rate averaged viscous number $J'$. The rheology and microstructure of dense suspensions are studied by systematically varying the strain magnitude $\gamma_0$ and $J'$ using numerical simulations. We study both suspensions composed of frictional ($\mu_p=0.4$) or frictionless ($\mu_p=0$) particles and find that the critical values, as $J' \to 0$, of both the complex macroscopic friction and the number of total and sliding contacts decrease with decreasing $\gamma_0$.
For suspensions composed of frictional particles, we also find that the critical (\emph{i.e.,}~the shear jamming) packing fraction $\phi_c$ increase with decreasing $\gamma_0$. 
In both cases, frictional and frictionless, we find that the rheological response approaching the shear jamming turns from a viscous to an elastic response as $\gamma_0$ is lowered below $\sim0.33$.

\end{abstract}

\maketitle

%%%MAIN TEXT%%%%
\section{Introduction}
Oscillatory shear measurements are widely used when studying mechanical properties (such as viscosity and elasticity) of soft materials and complex fluids such as suspensions, emulsions, polymer melts[\onlinecite{ferry1980viscoelastic,dealy2012melt,marenne2017nonlinear}]. In a typical oscillatory shear measurement, the material is subject to an oscillating shear-rate, \textit{i.e.}~$\dot\gamma=\dot\gamma_0\cos(\omega t)$, or equivalently an oscillating strain, \textit{i.e.}~$\gamma = \gamma_0\sin(\omega t)$, of frequency $\frac{\omega}{2\pi}$ and with magnitudes equal to $\dot \gamma_0=\gamma_0 \omega$ during which the mechanical response is monitored as a function of time. For a purely viscous material, the mechanical response will always be in-phase with the applied shear-rate; while for a purely elastic material the mechanical response will be in-phase with the applied strain (and out-of-phase with the applied shear-rate). In many cases, the tested materials are viscoelastic, which means that the mechanical response contains both in-phase and out-of-phase parts with respect to the shear-rate (or strain). The viscoelastic response is usually quantified by the elastic storage modulus $G'$ and the viscous loss modulus $G''$, or equivalently by the imaginary part of the complex viscosity $\eta''$ and the real part of the complex viscosity $\eta'$. The former one accounts for the mechanical response that is in-phase with the strain, and the latter one accounts for the part that is in-phase with the shear-rate[\onlinecite{hyun2011review}].\\
\newline
The rheology of non-Brownian non-inertial dense suspensions composed of hard particles at steady shear (\emph{i.e.,}~at a constant shear-rate $\dot\gamma$) can be characterised by one single dimensionless parameter: the viscous number $J=\eta_f\dot\gamma/P$[\onlinecite{Boyer11}],
 where $P$ is the pressure (normal stress) and $\eta_f$ the interstitial fluid viscosity. This characterisation assumes that the interstitial fluid is viscous enough such that particle inertia can be ignored. Both the packing fraction $\phi$ and the macroscopic friction (or stress ratio) $\mu=\sigma/P$, $\sigma$ being the shear stress, can then described by two single-variable constitutive relationships: $\phi=\phi(J)$ and $\mu=\mu(J)$[\onlinecite{Boyer11,Trulsson12,Trulsson17}]. For non-Brownian dense suspensions, one is typically interested in how the rescaled viscosity $\eta/\eta_f(=\mu/J)$ varies with the packing fraction of the solid material $\phi$ (which is experimentally more accessible than $J$). The divergence of the viscosity from below the shear jamming packing fraction $\phi_c$ is found empirically (or semi-empirically) to follow a power-law as $\eta/\eta_f\sim(\phi_c-\phi)^{-n}$, where $n$ is a positive exponent and usually close to two[\onlinecite{Andreotti12,Boyer11}]. 
 However, the values of $\phi_c$ do depend on several factors, for example, the shape of the particles[\onlinecite{Trulsson18}], the type of interactions between the particles[\onlinecite{Irani14}], the dimensionality of the system[\onlinecite{Olsson19}], and the particle-particle friction[\onlinecite{Silbert10,Trulsson17}]. A key result from above analysis is that the viscosity of a non-Brownian non-inertial suspensions is shear-rate-independent for a given fixed packing fraction. \\
%Dense suspensions that are subjected to oscillatory shear flows do not always respect the Cox-Merz rule and, hence, the steady shear rheology can not always be inferred from the corresponding oscillatory shear rheology or vice versus \onlinecite{Guazzelli18}. Despite that, the literature is surprisingly scarce when it comes to studying oscillatory shear rheology of dense suspensions and we have only a limited knowledge about dense suspensions viscoelastic properties. \\
\newline
Things do, in general, become more complicated under unsteady conditions, so also for suspensions subjected to oscillatory shear flows [\onlinecite{Erwin10}].
 Oscillatory shear introduces both a frequency and a magnitude into the rheological response.
Traditionally and depending on the strain magnitude, the oscillatory shear flows are usually characterised into a a Small Amplitude Oscillatory Shear (SAOS)-regime and a Large Amplitude Oscillatory Shear (LAOS)-regime. Such studies are usually carried out under constant packing fraction. In the SAOS-regime, all stresses show linear response, and one is effectively scanning a material's linear viscoelastic properties[\onlinecite{van1989linear,shikata1994viscoelastic}]. The LAOS-regime is characterised as the non-linear regime, where stresses deviate from simple linear responses[\onlinecite{hyun2011review,bricker2006oscillatory,lin2013short,park2011rheology,wang2016large,Ewoldt08}]. The non-linearity in LAOS is often measured using Fourier Transformation[\onlinecite{hyun2011review,wilhelm2002fourier,Ewoldt08}] or Chebyshev polynomial decomposition[\onlinecite{wang2016large,cho2005geometrical}] in frequencies, and is far from trivial to analys.\\ \newline
LAOS has been used to study complex behaviors of various suspensions. For example, it is used to probe microstructure and rheology of thixotropic suspensions composed of colloidal particles [\onlinecite{min2014microstructure,armstrong2016dynamic}].
LAOS can, furthermore, be used to yield glassy hard sphere colloidal suspensions, where the suspension is doing a transition from a Brownian to an effectively non-Brownian suspension [\onlinecite{Koumakis13}], characterised by a high P\'eclet number. In our work, we deal with an idealised non-inertial non-Brownian suspensions, corresponding to having an infinite P\'eclet number,  and due to the fact that the rheology is rate-independent, the final quantities will have zero frequency dependencies. This simplifies the analysis considerably as one only need to account for the magnitude.  \\
%At steady shear (\emph{i.e.}~at a constant shear-rate $\dot\gamma$), the rheology of dense suspensions of hard particles can essentially be characterised by one single dimensionless parameter: the viscous number $J=\eta_f\dot\gamma/P$,\onlinecite{Boyer11},
%where $P$ is the pressure (normal stress) and $\eta_f$ the interstitial fluid viscosity, and  assuming that the interstitial fluid is viscous enough such that particle inertia can be ignored. Both the packing fraction $\phi$ and the macroscopic friction (or stress ratio) $\mu=\sigma/P$, where $\sigma$ is the shear stress, can then described by two single-variable constitutive relationships: $\phi=\phi(J)$ and $\mu=\mu(J)$.\onlinecite{Boyer11,Trulsson12,Trulsson17} For dense suspensions one is typical interested in how the rescaled viscosity $\eta/\eta_f(=\mu/J)$ varies with packing fraction of the solid material $\phi$. The divergence of the viscosity from below the shear jamming packing fraction $\phi_c$ is found empirically (or semi-empirically) to follow a power-law as $\eta/\eta_f\sim(\phi_c-\phi)^{-n}$, where $n$ is a positive exponent and usually close to two\onlinecite{Andreotti12,Boyer11}. 
%The values of $\phi_c$ do, however, depend on several various factors, for example shape of the particles\onlinecite{Trulsson18}, the type of interactions between the particles\onlinecite{Irani14}, the dimensionality of the system\onlinecite{Olsson19}, and the particle-particle friction\onlinecite{Silbert10,Trulsson17}.\\
\newline
Besides studying the rheology of dense suspensions in oscillatory shear, it has found to be insightful to look at the microstructure, \emph{e.g.,}~how the number of particle contacts evolves [\onlinecite{Mohan13}]. It has found that if one instantaneously reverses the direction of the shear, the viscosity of the suspension will first drop and then gradually increase back to its steady-shear value[\onlinecite{blanc2011local,Lin15,peters2016rheology}]. Such behaviour is explained by the breakage and re-establishment of the contacts between the particles in a suspension. Oscillatory shear flows at large strain magnitudes can considered as a series of shear reversals from a steady shear[\onlinecite{Guazzelli18}]. As the strain magnitude is lowered, the particles in the suspensions cannot fully restore the same microstructure as found at steady shear[\onlinecite{Ness17}], yielding a lower viscosity. It has been shown that if the strain magnitude is small enough, the trajectories of the particles in the suspensions can become fully reversible, indicating that the suspensions reach an ``absorbing'' state where the particles self-organised to avoid each other[\onlinecite{corte2008random,metzger2012clouds,pine2005chaos,Ness18}].\\
%One can note that the frequency does not play if a suspension is in its linear response regime. \\
\newline
Moreover, some recent studies have found that by applying an oscillatory shear perpendicular to a primary stationary shear, one can reduce the viscosity of the suspension[\onlinecite{Lin16,Ness17}] and, in some cases, even ``unjam'' suspensions that have been jammed by steady shear[\onlinecite{Ness18}]. A possible explanation of this is that the perpendicular oscillation tilts and eventually breaks the force chains in the suspensions[\onlinecite{Lin16}]. In our previous work[\onlinecite{Dong2020-1}], we found that for a two-dimensional system, an oscillatory shear parallel to the primary can also decrease the viscosity of a dense suspension, illustrating that the viscosity reduction is primarily due to a general restructuring of the microstructure. We further showed that the shear-jamming packing fractions were pushed to higher values for small oscillatory strains for suspensions composed of frictional particles but not for frictionless.  \\ \newline
In this work, we continue our previous study[\onlinecite{Dong2020-1}] and present additional numerical results on the rheology on non-Brownian dense suspensions of hard particles immersed in a Newtonian fluid under oscillatory shear in a pressure-controlled set-up.  Our new results show that dense suspensions have a viscoelastic response if the oscillatory strain magnitudes are small enough (typically below 0.33). At the same conditions, we find that the critical values of the complex macroscopic friction $|\mu^{*}|_c$ and the number of contacts $Z_c$ (as $\dot \gamma\to 0$ or $P\to \infty$) saturate to values significantly below the steady shear ones. These changes are attributed to an altered microstructure at small strain magnitudes compared to steady shear. 

\section{Simulation Methods}
We study dense suspensions in two dimensions consisting of $\sim1000$ polydisperse circular discs, with an average diameter equal to $d$, using a discrete element method. The particles of the suspensions are confined between two rough walls, which both are exposed to a constant imposed pressure $P^{ext}$ in the normal direction (now denoted the $y$-direction) and a relative oscillatory velocity difference between them along the $x$-axis (\emph{i.e.}~the tangential direction), the latter resulting in a macroscopic shear-rate $\dot\gamma(t)=\dot\gamma_0\cos(\omega t)$, where $\dot\gamma_0$ is the magnitude of the oscillatory shear and $\omega/2\pi$ the frequency of the oscillations. 
The strain is then given as $\gamma(t) = \gamma_0\sin(\omega t)$ with the strain magnitude of the oscillations $\gamma_0=\dot\gamma_0/\omega$ (corresponding to the maximum strain in one direction).
The walls are constructed by fusing the same kind of discs as compose the flowing discs. 
Interactions between discs are described by harmonic springs[\onlinecite{Cruz05}]
\begin{equation}\label{eq1}
\mathbf{f}_{ij} = k_n\delta_n^{ij}\mathbf{n}^{ij}+k_t\delta_i^{ij}\mathbf{t}^{ij},
\end{equation}
where $k_n$ is the normal spring constant, $k_t=k_n/2$ the tangential spring constant, $\delta_n^{ij}$ the normal overlap, and $\delta_t^{ij}$ the tangential displacement. The ratio between the normal spring constant and the imposed pressure $k_n/P^{\mathrm{ext}}$ is kept equal to $3\cdot10^4$ to ensure a near-hard disc condition. Each tangential force is restricted by a Coulomb friction criterion, $|f_t^{ij}|\leq\mu_p|f_n^{ij}|$, where the particle-particle friction coefficient was chosen to be either $\mu_p=0.4$ (frictional) or $0$ (frictionless). The discs are also subject to both viscous drag and torque, given as
\begin{eqnarray}\label{eq2}
\mathbf{f}_i^v=\frac{3\pi\eta_f}{1-\phi_0}[\mathbf{u}^f(y)-\mathbf{u}_i], \\
\label{eq3}
\tau^v_i=\frac{\pi\eta_fd^2}{1-\phi_0}[\omega^f-\omega_i],
\end{eqnarray}
with $\phi_0=0.76$ being a typical packing fraction, and where $\mathbf{u}_i$ and $\omega_i$ are the translational and angular velocities of disc $i$, $\mathbf{u}^f(y)=(\dot\gamma y, 0)$ the interstitial fluid velocity where $y$ is the $y-$coordinate of disc $i$, and $\omega^f=\dot\gamma/2$ the fluid angular velocity (vorticity). Both simulations with and without lubrication forces and torques were considered. In the case of simulations with lubrication forces, the disc dynamics is described by Newtonian dynamics, where we set the maximum Stokes number $St = \rho\dot\gamma_0 d^2/\eta_f$ smaller than $0.4$ so that there are negligible inertial effects. The results of simulations with lubrication force are presented in section I in the SI, where we show that including lubrication forces gives similar results compared to the results without lubrication forces. In the case of simulations without lubrication forces,
the disc dynamics is strictly overdamped and given by a force and a torque balance
\begin{eqnarray}\label{eq4}
\mathbf{f}_i^{ext}+\mathbf{f}_i^v=-\sum_j\mathbf{f}_{ij}, \\
\label{eq5}
\tau_i^{ext}+\tau_i^{v}=-\sum_j\tau_{ij},
\end{eqnarray}
on each disc and where $\mathbf{f}_i^{ext}$, $\mathbf{f}_i^v$ and $\mathbf{f}_{ij}$ are the external, viscous and contact forces respectively and $\tau_i^{ext}$, $\tau_i^{v}$ and $\tau_{ij}$ the corresponding torques. In the overdamped simulations, we used a time-step equal to a tenth of the characteristic time-scale $t_0 = \frac{3\pi\eta_f}{(1-\phi_0)k_n}$.\\
\newline
For each simulation, we fix the confining pressure $P^{ext}$, the magnitude of oscillatory shear $\dot\gamma_0$ and the oscillation frequency $\omega$.
The suspensions are ``pre-equilibrated'' before any properties are measured, such as to avoid drift and transient behaviours. 
%Averages of each oscillation period do not have any clearly drift but only fluctuate around certain mean values. 
We accumulated statistics over a minimum of 10 absolute strains ($\int |\dot\gamma|dt \geq 10$) and at least one full oscillation period (for the largest $\gamma_0$).
All reported quantities are measured by excluding the first five layers close to each wall to eliminate boundary effects. 
%We sample in total 60 measurements (evenly distributed in time) per oscillations.
 Sampling frequency was put equal to 60 measures (evenly distributed in time) per oscillation. In the following, we only report the particle contact and lubrification stresses, ignoring the stress coming from the interstitial fluid itself ($\sigma_{\rm fluid}=\eta_f \dot \gamma$).
 
\section{Suspensions Rheology}
For suspensions subjected to oscillatory shear, the shear stress $\sigma$ can consist of contributions from both elastic and viscous responses. In the linear response regime, the shear stress can be written as[\onlinecite{marenne2017nonlinear,ishima2020scaling}]
\begin{equation}\label{eq:sigma_tot}
\sigma = G'\gamma(t)+\frac{G''}{\omega}\dot\gamma(t)=\eta''\dot\gamma_0\sin(\omega t)+\eta'\dot\gamma_0\cos(\omega t),
\end{equation}
where $G'$ is the storage modulus, $G''$ is the loss modulus, $\eta'$ is the real part of the complex viscosity, and $\eta''$ is the imaginary part of the complex viscosity with $\eta^{*}=\eta'-i\eta''$ defined as the complex viscosity. The magnitude of the complex viscosity is given by $|\eta^{*}|=\sqrt{\eta'^2+\eta''^2}$. $\eta'$ and $\eta''$ can be calculated from the shear-rate averaged, or strain averaged shear stresses as[\onlinecite{Dong2020-1,otsuki2020shear}]
\begin{equation}
\eta' = \frac{\int_0^{2\pi/\omega}\sigma(t)\cos(\omega t) \,dt}{\dot\gamma_0\int_0^{2\pi/\omega}\cos^2(\omega t) \,dt}= \frac{\int_0^{2\pi/\omega}\sigma(t) \dot \gamma(t) \,dt}{\int_0^{2\pi/\omega} \dot \gamma(t)^2 \,dt}=G''/\omega,
\end{equation}
\begin{equation}
\eta'' = \frac{\int_0^{2\pi/\omega}\sigma(t)\sin(\omega t) \,dt}{\dot\gamma_0\int_0^{2\pi/\omega}\sin^2(\omega t) \,dt}=\frac{\int_0^{2\pi/\omega}\sigma(t) \gamma(t) \,dt}{\int_0^{2\pi/\omega} \gamma(t)^2 \,dt}=G'/\omega.
\end{equation}
Note that these quantities can always be calculated and defined even if one is not in a linear regime. %(see Eq.~\ref{eq:sigma_tot}).
Following our previous approach[\onlinecite{Dong2020-1}], we generalise Eqs.~7 and 8, applying it also to pressures/normal stresses (see the SI) and stress ratios
\begin{equation}
\mu' = \frac{\int_0^{2\pi/\omega} (\sigma/P)\dot\gamma(t) \,dt}{\int_0^{2\pi/\omega}|\dot\gamma(t)| \,dt},
\end{equation}
\begin{equation}
\mu'' = \frac{\int_0^{2\pi/\omega} (\sigma/P)\gamma(t) \,dt}{\int_0^{2\pi/\omega}|\gamma(t)| \, dt},
\end{equation}
where $\mu'$ is the viscous component, $\mu''$ is the elastic component of the complex macroscopic friction coefficient $\mu^{*}$, with the magnitude equal to $|\mu^{*}|=\sqrt{\mu'^2+\mu''^2}$.\\
%, and  is the magnitude of the complex macroscopic friction coefficient.\\
%Similar as Eq.~\ref{eq:sigma_tot}, we can write an expression for pressure $P$ of the suspensions. The difference is that the pressure do not change sign with $\dot\gamma$ or $\gamma$ and thus only depending on the magnitude of $\dot\gamma$ and $\gamma$,
%\begin{equation}\label{eq:P_tot}
%P = G_P'|\gamma(t)|+\frac{G_P''}{\omega}|\dot\gamma(t)|=\eta_P''|\dot\gamma_0\sin(\omega t)|+\eta_P'|\dot\gamma_0\cos(\omega t)|,
%\end{equation}
%where
%\begin{equation}
%\eta_P' = \frac{\int_0^{2\pi/\omega}P(t)|\cos(\omega t)|dt}{\dot\gamma_0\int_0^{2\pi/\omega}\cos^2(\omega t)dt},
%\end{equation}
%\begin{equation}
%\eta_P'' = \frac{\int_0^{2\pi/\omega}P(t)|\sin(\omega t)|dt}{\dot\gamma_0\int_0^{2\pi/\omega}\sin^2(\omega t)dt}.
%\end{equation}
%We define $\mu'=\eta'/\eta_P'$ the macroscopic friction coefficient from viscous response, $\mu''=\eta''/\eta_P''$ the macroscopic friction coefficient from elastic response, and $|\mu^{*}|=\sqrt{\mu'^2+\mu''^{2}}$. \\
\newline
A shear-rate averaged viscous number $J'$ is defined in the same way as in [\onlinecite{Dong2020-1}],
\begin{equation}
J' = \frac{\eta_f \int_0^{2\pi/\omega} (\dot\gamma/P)\dot\gamma \,dt}{\int_0^{2\pi/\omega}|\dot\gamma| \,dt}.
\end{equation}
One can equivalent define $J''$, but can easily be seen to be equal to zero due to the $\dot \gamma \gamma$-term.
Other properties such as the packing fraction $\phi$ and the number of contacts $Z$ are generally less sensitive to the averaging process, being that the shear-rate-weighted-average, the strain-weighted-average, or the direct time-average (see the SI for a comparison). We, therefore, calculate the average of $\phi$ and $Z$ as 
\begin{equation}\label{eq6}
A = \frac{\int_0^{2\pi /\omega}A(t)(|\dot\gamma(t)|+|\omega\gamma(t)|)\,dt}{\int_0^{2\pi /\omega}(|\dot\gamma(t)|+|\omega\gamma(t)|)\,dt},
\end{equation}
where $A$ is either $\phi$ or $Z$. As $\int_0^{2\pi /\omega}(|\dot\gamma(t)|+|\omega\gamma(t)|)\,dt=2$, this is in principal the arithmetic mean between the strain and shear-rate averaged quantities.
%The suspensions are pre-sheared before any properties are measured and the measurement is usually over either $10$ oscillation periods (when $\mathcal{G}$ is small) or an absolute strain $\int |\dot\gamma|dt=10$ (when $\mathcal{G}$ is large). Values of $J_{|\dot\gamma|}$ is varied by changing $\dot\gamma_1$ while keeping $P$ and $\mathcal{G}=\dot\gamma_1/\omega$ constant. Time series of all reported quantities are measured with first $5$ layers close to each walls excluded to eliminate boundary effects.
\section{Results and Discussions}
\subsection{VISCOELASTIC RESPONSE}
\subsubsection{Time series and Lissajous curves}
\begin{figure*}[!htbp]
\centering
\includegraphics[width=0.45\textwidth]{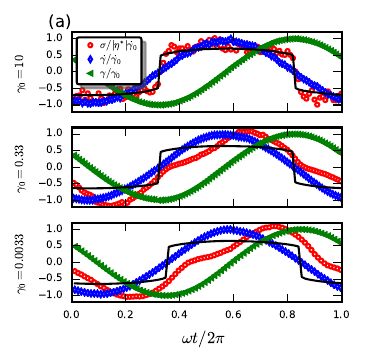}
\includegraphics[width=0.45\textwidth]{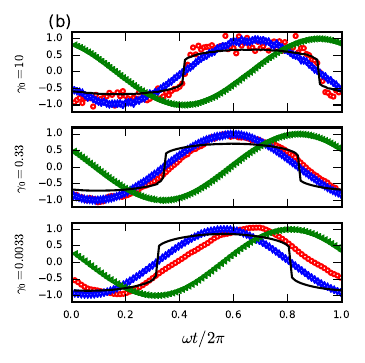}
  \caption{Time series of rescaled shear stress $\sigma/|\eta^{*}|\dot\gamma_0$ (red circles) where $|\eta^{*}|$ is the magnitude of complex viscosity for the corresponding case, shear-rate $\dot\gamma/\dot\gamma_0$ (blue diamonds), strain $\gamma/\gamma_0$ (green triangles) for (a)~frictional suspensions ($\mu_p=0.4$) and (b)~frictionless suspensions ($\mu_p=0$) at $J'\simeq 3\cdot 10^{-3}$  and three different $\gamma_0$ ($10$, $0.33$, and $0.0033$) as indicated in the labels to the left of the figures. The values of $|\eta^{*}|/\eta_f$ are $139\, (\gamma_0=10)$, $194\, (\gamma_0=0.33)$, and $180\, (\gamma_0=0.0033$) in (a), and $99\, (\gamma_0=10)$, $96\, (\gamma_0=0.33)$, and $76\, (\gamma_0=0.0033)$ in (b).
The black lines are analytical prediction corresponding to a purely viscous stress response.}
  \label{fgr:time-series}
\end{figure*}

\begin{figure*}[!htbp]
\centering
\includegraphics[width=0.45\textwidth]{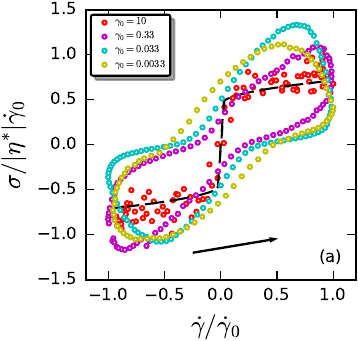}
\includegraphics[width=0.45\textwidth]{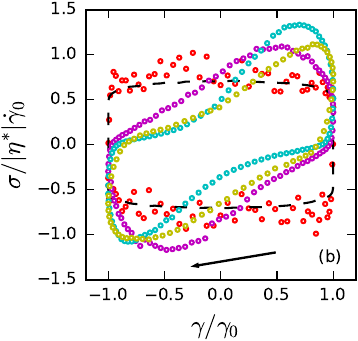}
  \caption{Lissajous curves of stress for frictional suspensions ($\mu_p=0.4$) at $J' \simeq 3\cdot 10^{-3}$ and various strain magnitude $\gamma_0$ as indicated in the legends, as functions of (a)~shear-rate $\dot\gamma/\dot\gamma_0$ and (b)~strain $\gamma/\gamma_0$.  
  The shear stresses $\sigma$ are normalised as in Fig.~\ref{fgr:time-series}, with $|\eta^{*}|/\eta_f$-values equal to $139\, (\gamma_0=10)$, $194\, (\gamma_0=0.33)$, $238\,  (\gamma_0=0.033)$ and $180\, (\gamma_0=0.0033)$.
  The black dashed lines are plots of the analytical predictions, and the arrows indicate the directions.}
  \label{fgr:Lissajous}
\end{figure*}

\begin{figure*}[!htbp]
\centering
\includegraphics[width=0.45\textwidth]{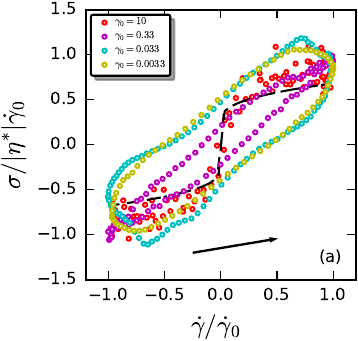}
\includegraphics[width=0.45\textwidth]{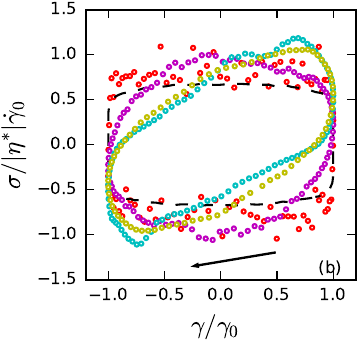}
  \caption{As in Fig.\ref{fgr:Lissajous} but for frictionless suspensions ($\mu_p=0$). The values of $|\eta^{*}|/\eta_f$ are $99\, (\gamma_0=10)$, $96\, (\gamma_0=0.33)$, $97\, (\gamma_0=0.033)$ and $76\, (\gamma_0=0.0033)$.}
  \label{fgr:Lissajous-nf}
\end{figure*}

In Fig.~\ref{fgr:time-series}, we show some examples of time series of the rescaled stress $\sigma/|\eta^{*}|\dot\gamma_0$ compared to the rescaled shear-rate $\dot\gamma/\dot\gamma_0$ and the strain $\gamma/\gamma_0$ at $J'\simeq 3\cdot 10^{-3}$ for suspensions composed of (a)~frictional or (b)~frictionless particles at three different strain magnitudes. %The shear-rate magnitude $\dot \gamma_0\sim 3.5\sim 10^{-3}$.
The black lines in the sub-figures are all analytical predictions of how the stress will respond if its corresponding steady-shear rheology could describe the suspensions at each given time (\emph{i.e.}~corresponding to an instantaneous relaxation of the microstructure and stresses upon changes in shear-rate).
The analytical predictions are given by $\sigma=\mu(J)P$ and $\mu(J)=\mu_c+a_{\mu}J^{n_{\mu}}$, where $\mu_c=0.28$, $a_{\mu}=2.5$, $n_{\mu}=0.5$ for the frictional case and $\mu_c=0.09$, $a_{\mu}=1.3$, $n_{\mu}=0.38$ for the frictionless case[\onlinecite{Dong2020,Dong2020-1}]. In these predictions, we have accounted for that $\phi$ varies with a varying $\dot \gamma$ at constant pressure or equivalent to an instantaneous $J$. 
The average packing fractions of the dense suspensions at oscillatory shear flows are generally found to increase as $\gamma_0$ decreases (see SI). For the specific cases in Fig.~\ref{fgr:time-series} we find the couples $[\gamma_0, \phi]$: $[10,0.764]$, $[0.33,0.800]$, and $[0.0033, 0.835]$ in  (a) and $[10,0.817]$, $[0.33,0.823]$, and $[0.0033, 0.839]$ in (b). \\
\newline
For $\gamma_0=10$ the stress curves, both for the frictional and frictionless case, follow our analytic predictions with an excellent agreement (see Fig.~\ref{fgr:time-series}), illustrating that at large strain rates, the oscillatory rheology can be indeed inferred from the steady shear one (\emph{i.e.,}~respecting the Cox-Merz rule). 
At first sight, the stress response seems non-linear (\emph{i.e.,}~not being a simple linear combination of the strain and strain-rate curves), which could be considered as being in the non-linear LAOS-regime. However, we find here that the simulation results well-follow our analytical prediction, which is a clear sign that at $\gamma_0=10$ the suspension behaves as if under steady shear at each given time. The seemingly non-linearity in the stress response is instead caused by the varying packing fraction in our pressure controlled set-up (see the SI for the corresponding time-series of $\phi$) and is well captured by our analytical prediction.
%It should be noted that despite the large strain magnitude we are still at the low shear-rate regime here with the shear-rate magnitude $\gamma_0\sim 3.5\sim 10^{-3}$. 
Both experimental and numerical shear reversal results at constant volume [\onlinecite{blanc2011local,peters2016rheology}] have shown that one recovers the steady-shear microstructure after a shear-reversal and a finite strain, with the viscosity saturating to the steady-shear value. Our results for large strain amplitudes imply  that the same occurs under constant pressure.
%At large amplitudes, a suspension is able to recover its steady-shear microstructure under shear-reversals and hence shows the same rheological behavior as if under steady shear after a finite strain. Such observation is in agreement with}
%Although the stress response seems to be non-linear (\emph{i.e.}~not being a simple linear combination of the strain and strain-rate curves) it follows our analytical prediction perfectly and the non-linearities are, hence, caused by the varying packing fraction in our pressure controlled set-up (see SI for the corresponding time-series in $\phi$). 
It should be further noted that our steady-state predictions only include viscous responses (or equivalent loss moduli) in the description of the stresses; hence an agreement with analytic predictions implies a pure (linear) viscous response. In real systems, one usually finds non-linearities at large amplitudes, \emph{i.e.}~large shear-rates, due to, \emph{e.g.,}~inertial effects and/or particle deformations. These effects have not been considered in the current model, and, hence, the Cox-Merz rules arises naturally from the linearity of the system.
As $\gamma_0$ decreases, the stress curves, in both cases (frictional and frictionless), start to deviate from the analytical predictions, implying a viscoelastic response.\\
\newline
An alternative way to represent and illustrate the viscoelastic response is by Lissajous curves. 
In Fig.~\ref{fgr:Lissajous} and \ref{fgr:Lissajous-nf}, we show Lissajous curves for suspensions composed of frictional or frictionless particles respectively at various strain magnitudes as indicated by the legend. 
The Lissajous curves show the rescaled shear stresses as functions of (a)~shear-rate $\dot\gamma/\dot\gamma_0$ and (b)~strain $\gamma/\gamma_0$. The arrows indicate the directions of the curves. As before, the black dashed lines are analytical predictions assuming that the stress responses are linear, viscous, instantaneous, and described by their corresponding steady-shear rheology. Comparing Fig.~\ref{fgr:Lissajous} with \ref{fgr:Lissajous-nf}, we do not see much difference in the Lissajous curves between frictional and frictionless suspensions at the same $\gamma_0$. In both cases, the Lissajous curves with $\gamma_0=10$ are in good agreement with the analytical predictions, indicating that the behaviours of the suspensions still follow steady-shear rheology at large strain magnitudes and hence respect the Cox-Merz rule even at pressure imposed conditions. As $\gamma_0$ decreases, the Lissajous curves deviate from the steady-shear predictions.
At $\gamma_0 < 0.1$, which is usually associated with the SAOS-regime, one would expect to see ellipses in the Lissajous curve, being a indicator of linear viscoelastic response. In the cases here, however, we observe twisted ellipses.  It is hard to say if such a shape is evidence of a non-linear response or a result of a varying packing fraction (still with linear viscoelasticity).  or a result from a varying packing fraction (still with linear viscoelasticity). \\ 
\newline
%and have shapes of twisted ellipses, which an indication of a viscoelastic and/or non-linear response. \newline
In general, it is hard to directly map our results into the framework of SAOS and LAOS due to the pressure imposed set-ups we use here. To do so,
% interpret non-linearities and viscoelasticity from just looking at the stress curves for pressure imposed rheometers, as 
ne needs to know how packing fractions vary and their consequences for both the viscous and elastic response. Currently, we lack information about the latter response. In the following, we will try to circumvent this difficulty by looking at the complex, real, and imaginary viscosities and our generalisation of a complex macroscopic friction coefficient. %In the frictional case, the phase shift can be clearly seen already at $\gamma_0=0.33$ while in the frictionless case the shift is still minor at this $\gamma_0$. In addition, one observes irregularity in the stress curves at small $\gamma_0$ which is usually attribute to non-linear response if the suspensions are sheared under constant packing fractions. However, the suspensions are sheared under constant imposed pressure here. The irregularity in the stress curves are more likely a result of the fluctuation in the packing fractions. See SI for plots of time series of packing fractions.  
%When plotted as functions of shear-rate, the Lissajous curve will be a straight if the stress response is pure linear and a circle if the response is pure elastic, and vice versa if plotted as functions of strain. In both cases, ellipses indicate linear viscoelastic response. Shapes that depart from these three conditions imply non-linear stress responses. Comparing Fig.~\ref{fgr:Lissajous} with \ref{fgr:Lissajous-nf}, we can see that the Lissajous curves for frictional and frictionless suspensions at the same $\gamma_0$ are similar. At large strain magnitude $\gamma_0=10$ and $3.3$, the Lissajous curves suggest non-linear viscous response, with S-shaped lines in the $\sigma-\dot\gamma$ plot and square shape in the $\sigma-\gamma$ plot. At intermediate to small strain magnitude $\gamma_0$ between $0.0033$ and $1$, the shapes of the Lissajous curves become ellipse-like with some extends of deform indicating viscoelastic response with some non-linearity.
\subsubsection{Total response}
\begin{figure*}[!htbp]
\centering
\includegraphics[width=0.45\textwidth]{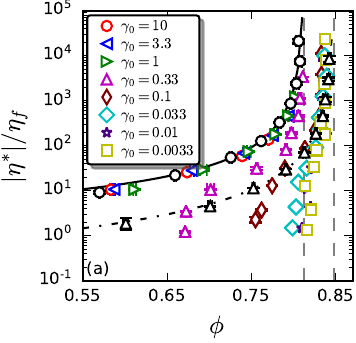}
\includegraphics[width=0.45\textwidth]{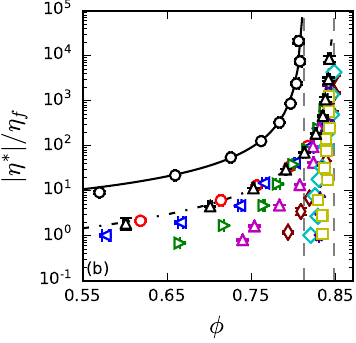}
  \caption{Rescaled complex viscosities $|\eta^{*}|/\eta_f$ as a function of the packing fraction $\phi$ at various strain magnitude $\gamma_0$ as indicated in the legends for the suspensions composed of (a)~frictional ($\mu_p=0.4$) and (b)~frictionless ($\mu_p=0$) particles. The black circles and black triangles are viscosities under steady shear for frictional and frictionless cases, respectively. The black lines are constitutive laws for suspensions under steady shear. The constitutive laws are given as $\eta/\eta_f = a(\phi-\phi_c)^{-n}+a'(\phi-\phi_c)^{1-n}$, where $a=0.18$, $n=2$, $a'=1.9$ for the frictional case and $a=0.004$, $n=2.7$, $a'=0.18$ for the frictionless case[\onlinecite{Dong2020}]. $\phi_c$ is the shear jamming packing fraction for suspensions under steady shear as indicated by the grey vertical lines with $\phi_c^{\mathrm{f}}=0.812\pm0.002$ for frictional and $\phi_c^{\mathrm{nf}}=0.848\pm0.002$ for frictionless cases. }
  \label{fgr:eta}
\end{figure*}

\begin{figure*}[!htbp]
\centering
\includegraphics[width=0.33\textwidth]{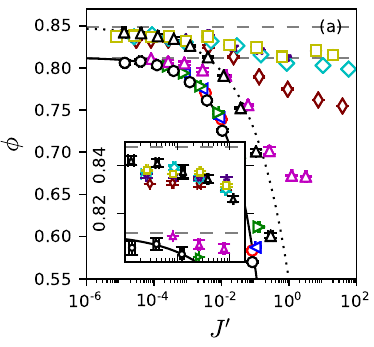}
\includegraphics[width=0.33\textwidth]{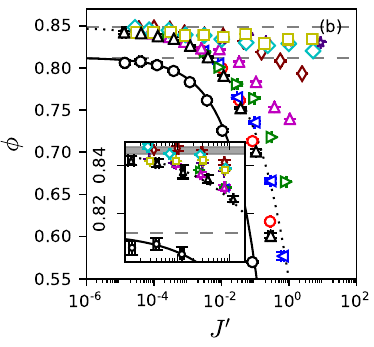}
\includegraphics[width=0.315\textwidth]{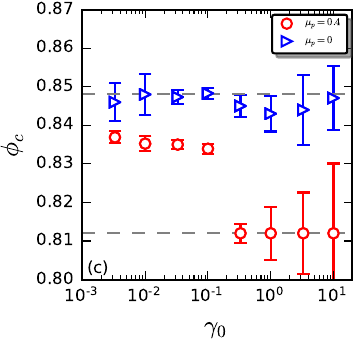}
  \caption{Packing fraction $\phi$ as a function of the viscous number $J'$ at various $\gamma_0$ for (a)~frictional and (b)~frictionless suspensions. The colours and symbols are the same as in Fig.~\ref{fgr:eta}. The black symbols correspond to the steady-shear conditions, and the black lines are plots of the constitutive laws for the steady-shear cases. The constitutive laws are $\phi=\phi_c-a_{\phi}J^{n_\phi}$, where $a_{\phi}=0.79$, $n_{\phi}=0.5$ for the frictional case and $a_{\phi}=0.3$, $n_{\phi}=0.38$ for the frictionless case[\onlinecite{Dong2020}]. The grey dashed horizontal lines indicate the values of $\phi_c$ for suspensions under steady shear (the values of $\phi_c$ are given in the caption of Fig.~\ref{fgr:eta}). The insets are zoomed-in figures of the main figures close to jamming. In (c) jamming packing fraction $\phi_c$ as a function of $\gamma_0$ for both the frictional and frictionless suspensions, the grey dashed horizontal lines are again indications of the steady-shear values of $\phi_c$.}
  \label{fgr:phi}
\end{figure*}

\begin{figure*}[!htbp]
\centering
\includegraphics[width=0.35\textwidth]{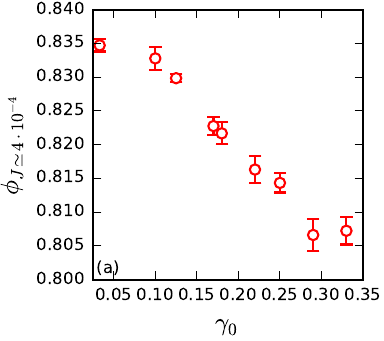}
\includegraphics[width=0.33\textwidth]{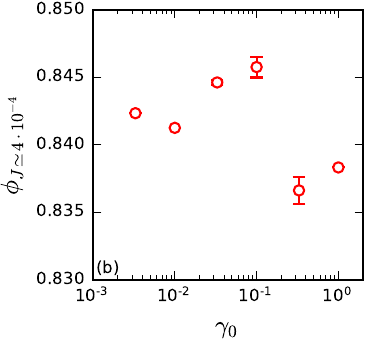}
  \caption{Packing fraction $\phi$ as a function of the strain magnitude $\gamma_0$ at $J'\simeq 4\cdot 10^{-4}$ for suspensions composed of (a)~frictional particles and (b)~frictionless particles.}
  \label{fgr:phi-gamma}
\end{figure*}

\begin{figure*}[!htbp]
\centering
\includegraphics[width=0.33\textwidth]{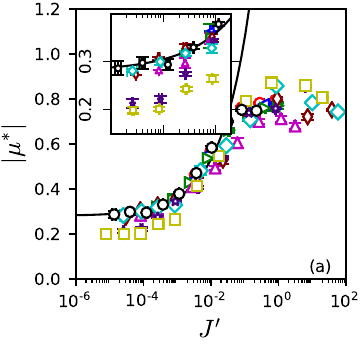}
\includegraphics[width=0.33\textwidth]{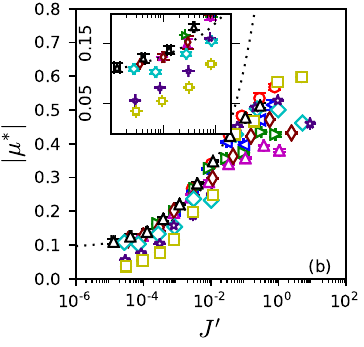}
\includegraphics[width=0.325\textwidth]{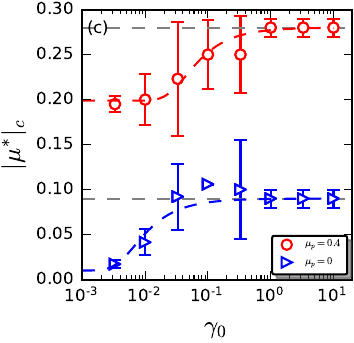}
  \caption{Complex macroscopic friction $|\mu^{*}|$ as a function of the viscous number $J'$ at various $\gamma_0$ for (a)~frictional and (b)~frictionless suspensions. The colours and symbols follow the same styles as in Fig.~\ref{fgr:eta}, with zoomed-in figures close to $J'\to 0$ as insets. The black symbols are data for the suspensions under steady shear, and the black lines are plots of the constitutive laws for the steady-shear cases. The constitutive laws are $\mu=\mu_c+a_\mu J^{n_{\mu}}$. The values of the parameters have been given in section~IV.A.1[\onlinecite{Dong2020}]. In (c) $|\mu^{*}|_c$ as a function of $\gamma_0$; the coloured dashed lines are best fits of the phenomenological function $|\mu^{*}|_c=\mu_c^{\rm{SS}}[1-k_1\tanh(k_2/\gamma_0)]$, where $\mu_c^{\rm{SS}}$ is the values of suspensions under steady shear with $\mu_c^{\mathrm{f,SS}}=0.28$ for the frictional suspensions and $\mu_c^{\mathrm{nf,SS}}=0.09$ for the frictionless suspensions; $k_1$ and $k_2$ are two fitted parameters, the values are given in text. The grey dashed horizontal lines indicate the values of the corresponding $\mu_c^{\mathrm{SS}}$. }
  \label{fgr:mu} 
\end{figure*}

Fig.~\ref{fgr:eta} shows complex viscosities as functions of packing fraction for the suspensions composed of (a)~frictional and (b)~frictionless particles. At large strain magnitudes $\gamma_0=10$ and $\gamma_0=3.3$, the viscosities have the same packing fraction dependencies as in their corresponding steady-shear cases[\onlinecite{Dong2020}] (indicated by black symbols and lines).
Above shows, together with the time-series of the stresses, that steady-state can be used to understand the oscillatory rheology at large oscillatory strains and vice versa, \emph{i.e.,}~that the Cox-Merz rule is applicable in these linear models.
%Hence, both the  and complex viscosity show that steady state can be used to understand the oscillatory rheology at large oscillatory strains. 
As $\gamma_0$ becomes smaller, the complex viscosities start to deviate from the steady-shear cases, as seen previously for the time-series and Lissajous plots of the corresponding cases, with a reduced complex viscosity as the strain amplitude is lowered. These trends correlate well with what we previously found when only considering the real part of the viscosity[\onlinecite{Dong2020-1}].  
The main difference between the frictional and the frictionless cases is that the jamming packing fraction $\phi_c$ for the frictional suspensions is shifted to a higher value at small strain magnitudes ($\gamma_0 \leq 0.1$). In contrast, for the frictionless suspensions $\phi_c$ does not show such apperent shift [\onlinecite{Ness18,Dong2020-1}].
%The main difference between the frictional and the frictionless cases is that the jamming packing fraction $\phi_c$ for the frictional suspensions shifted to a higher value at small strain magnitude $\gamma_0 \leq 0.1$ while for the frictionless suspensions $\phi_c$ do not show obvious shift. 
The effect of varying $\gamma_0$ on $\phi_c$ is better illustrated when studying $\phi$ as functions of shear-rate averaged viscous number $J'$, see Fig.~\ref{fgr:phi}. For fixed $\gamma_0$ values, 
all curves show a monotonically increasing packing fraction as $J'$ is decreased. 
Similar as in Fig.~\ref{fgr:eta}, we see that both $\gamma_0=10$ and $3.3$ curves collapse on top of the steady-shear curve.
As $\gamma_0$ is decreased, the packing fractions do, however, increase at large $J'$ values. 
As $J'$ approaches zero (and hence shear jamming with its corresponding critical packing fraction $\phi_c$) one sees a clear and almost discontinuous jump in the packing fraction curves when lowering the strain amplitude below $\gamma_0<0.33$ for the frictional case.  For $\gamma_0 > 0.1$ and frictional particles, $\phi_c$ coincides with the same value  as found in steady state, denoted by $\phi_c^{\mathrm{f,SS}}$, while for $\gamma_0 < 0.33$ $\phi_c$ is closer but still below the critical packing fraction found for frictionless particles in steady shear, denoted by $\phi_c^{\mathrm{nf,SS}}$. The inset of Fig.~\ref{fgr:phi}(a) shows a zoom-in close to shear jamming. For frictionless cases, a shift in $\phi_c$ is less apperant (see the inset of Fig.~\ref{fgr:phi}(b)). However, for the lowest studied $J'$-values we see a non-monotonic dependency of the packing fractions as a function of $\gamma_0$ and with the maximum packing fraction occurring at around $\gamma_0=0.1$ (see Fig.~\ref{fgr:phi-gamma}(b) for values of $\phi$ at $J'\simeq 4\cdot 10^{-4}$, one of our lowest $J'$ values). Nevertheless, all curves seem to have a packing fraction approximately equal or less to the critical packing fraction $\phi_c^{\mathrm{nf},SS} $for frictionless particles at steady shear. It should, however, be noted that the value of $\phi_c^{\mathrm{nf},SS}$ vary slightly ($\phi_c^{\mathrm{nf},SS} \in [0.845,0.848]$) depending on the range of $J'$ used when estimating it, this uncertainty is indicated by a shaded region in the inset of Fig.~\ref{fgr:phi}(b). Some of the packing fractions in the low $J'$ limit and at low $\gamma_0$ are all within this shaded area; hence we can not tell by certain if there is a shift or not in $\phi_c$. 
%At small $J$ and as $\phi$ approaches jamming packing fraction $\phi_c
%At large $J'$ values packing fractions increases as 
%As $\gamma_0$ becomes smaller, the packing fraction $\phi$ increases when $J$ is large for both the frictional and frictionless suspensions. 
We continue by estimating the $\phi_c$ values at different $\gamma_0$ for suspensions composed of either frictional or frictionless particles by fitting the data in the low $J'$-limit  (typically $J'<10^{-2}$) to $\phi=\phi_c+a_{\phi}(J')^{n_{\phi}}$, where $\phi_c$, $a_{\phi}$ and $n_{\phi}$ are all free parameters. Fig.~\ref{fgr:phi}(c) shows how the $\phi_c$ values vary as a function of $\gamma_0$.
For frictionless particles, we find $\phi_c$ to be the same or below $\phi_c^{\mathrm{nf,SS}}$. However, our standard error of the estimates are all on the order of the shifts; hence we currently lack the precision to actually conclude. For frictional particles, there is, however, a clear and seemingly discontinuous transition in the values of $\phi_c$ as $\gamma_0$ varies, from the values around $\phi_c^{\mathrm{f,SS}}$ at large strains to values around $\phi_c\sim0.835<\phi_c^{\mathrm{nf,SS}}$ as $\gamma_0$ drops below 0.33. To further investigate the nature of this transition in $\phi_c$ for the frictional particles, we run a few simulations at $J'\simeq 4\cdot 10^{-4}$ (\emph{i.e.}~close to shear jamming) and $\gamma_0\in[0.033, 0.33]$. The results are presented in Fig.~\ref{fgr:phi-gamma}(a), where we find a gradual increase in $\phi$ as $\gamma_0$ decreases, indicating that the transition of $\phi_c$ for the frictional suspensions is actually continuous in a narrow $\gamma_0$ range, starting at $\gamma_0=0.3$ and fully completed at $\gamma_0=0.05$. \\
\newline
Fig.~\ref{fgr:mu} shows how the complex friction coefficient $|\mu^{*}|$ varies as a function of $J'$ at various $\gamma_0$ for (a) frictional and (b) frictionless particles. In general, $|\mu^{*}|$ shows a similar trend as seen in steady shear, namely $|\mu^{*}|$ increases with $J'$ from a finite value $|\mu^{*}|_c$ at $J'=0$ (represented by the plateau at low $J'$ values in the lin-log representation).
At $\gamma_0=10$, $3.3$ and $1$, $|\mu^{*}|$ behaves approximately the same as for the steady-shear cases. At smaller $\gamma_0$, the $|\mu^{*}|$ values in the low $J'$-branch are all consistently lower than the corresponding steady-shear values. For larger $J'$ values, we could not see any clear trend (alternating between being both larger or smaller compared to the corresponding steady shear value as $\gamma_0$ is varied).
Values of  $|\mu^{*}|_c$  (\emph{i.e.} $|\mu^{*}|$ as $J'\to 0$) were extracted in the same way as was done for $\phi_c$, \emph{i.e.}~but fitting the data to $|\mu^{*}|=|\mu^{*}|_c+a_{\mu}(J')^{n_{\mu}}$ in the low $J'$-branch (typically  $J'<10^{-2}$).
%scatters around  higher values at large $J$ while lower values at small $J$, the latter is better shown in the insets of Fig.~\ref{fgr:mu}(a) and (b). 
%We extract the values of $|\mu^{*}|_c$ at different $\gamma_0$ by fitting the data that are close to the plateau to $|\mu^{*}|=|\mu^{*}|_c+a_{\mu}J^{n_{\mu}}$ ( $J<10^{-2}$) where $a_{\mu}$ and $n_{\mu}$ are fitting parameters. 
The results are plotted as a function of $\gamma_0$ in Fig.~\ref{fgr:mu}(c), where transitions can be seen for both the frictional and frictionless particles, from values equal to the steady-shear ones at large $\gamma_0$ to smaller values as $\gamma_0$ decreases. The transition can be described by a phenomenological hyperbolic function $|\mu^{*}|_c=\mu_c^{\rm{SS}}[1-k_1\tanh(k_2/\gamma_0)]$, where $\mu_c^{\rm{SS}}$ is the value of $\mu_c$ for the suspension at steady shear and $k_1$ and $k_2$ are two free parameters. $k_2$ signals at around which $\gamma_0$-value the transition occurs. For frictional particles we found $k_2=0.04\pm0.01$ and for frictionless particles $k_2=0.005\pm0.003$. $k_1$ instead signals the maximal decrease in $|\mu^{*}|_c$ due to oscillations, where $\mu_c^{\rm{SS}}(1-k_1)$ gives the value of $|\mu^{*}|_c$ as $\gamma_0\to 0$.  For frictional particles we found $k_1=0.29\pm0.03$ and for frictionless particles $k_1=0.88\pm0.25$, \emph{i.e.,}~a 30\% and 80-100\% decrease in $|\mu^{*}|_c$ respectively as $\gamma_0 \to 0$. 

%The values of $k_1$ and $k_2$ are given in Table~1.
%\begin{table}
%\begin{center}
%\caption{The values of $k_1$ and $k_2$ for the frictional and frictionless cases}
%\begin{tabular}{ |c|c|c| } 
% \hline
%  & frictional & frictionless \\ \hline 
% $k_1$ & $0.29\pm0.03$ & $0.88\pm0.25$ \\ 
% $k_2$ & $0.04\pm0.01$ & $0.005\pm0.003$ \\ 
 %\hline
%\end{tabular}
%\end{center}
%\end{table}
%where $k_1$ and $k_2$ are two fitting parameters, values of which are given in the caption of Fig.~\ref{fgr:mu}.
\subsubsection{Decomposition of the complex viscosity and macroscopic friction}
\begin{figure*}[!htbp]
\centering
\includegraphics[width=0.45\textwidth]{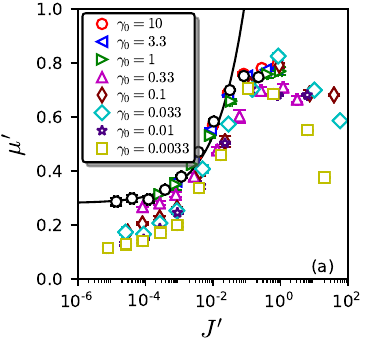}
\includegraphics[width=0.45\textwidth]{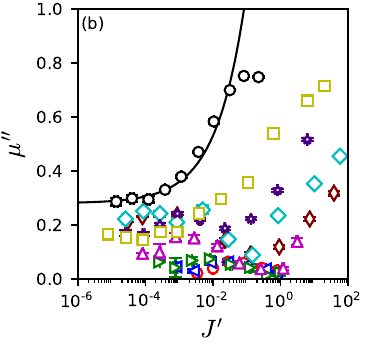}
  \caption{(a)~$\mu'$ and (b)~$\mu''$ as a function of $J'$ for the frictional suspensions at different $\gamma_0$ as indicated by the legends; the black circles are the values of $\mu$ for the frictional suspensions under steady shear, and the black lines are plots of the constitutive laws for the steady-shear case. The constitutive laws are the same as in Fig.~\ref{fgr:mu}(a).}
  \label{fgr:mu-ve}
\end{figure*}

\begin{figure*}[!htbp]
\centering
\includegraphics[width=0.45\textwidth]{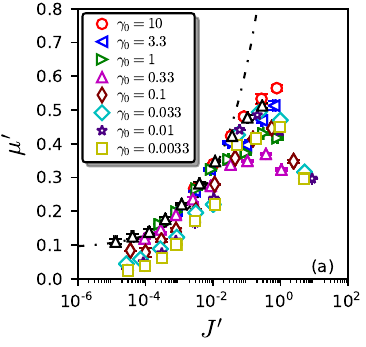}
\includegraphics[width=0.45\textwidth]{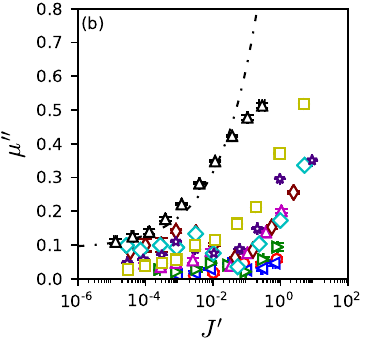}
  \caption{As in Fig.~\ref{fgr:mu-ve} but for frictionless particles, the constitutive laws are the same as in Fig.~\ref{fgr:mu}(b).}
  \label{fgr:mu-ve-nf}
\end{figure*}

\begin{figure*}[!htbp]
\centering
\includegraphics[width=0.45\textwidth]{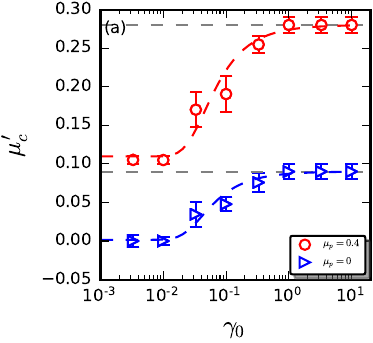}
\includegraphics[width=0.43\textwidth]{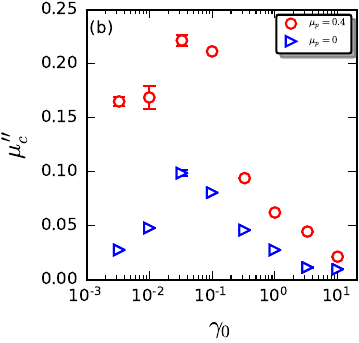}
  \caption{(a)~$\mu_c'$ and (b)~$\mu_c''$ as a function of $\gamma_0$ for suspensions composed of frictional or frictionless particles. Dashed grey horizontal lines in (a) show the corresponding values in steady shear. }
  \label{fgr:mupc}
\end{figure*}

\begin{figure*}[!htbp]
\centering
\includegraphics[width=0.45\textwidth]{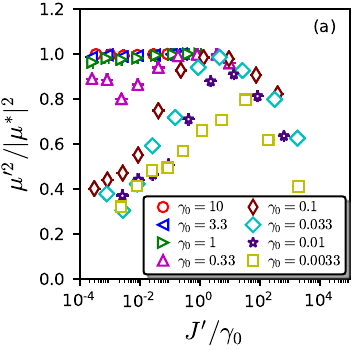}
\includegraphics[width=0.45\textwidth]{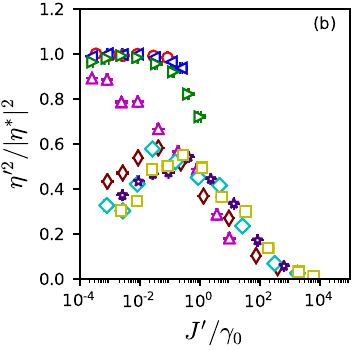}
  \caption{(a)~$\mu'^2/|\mu^{*}|^2$ and (b)~$\eta'^2/|\eta^{*}|^2$ as a function of $J'/\gamma_0$ for the suspensions composed of frictional particles at various $\gamma_0$ as indicated by the legend.}
  \label{fgr:decom}
\end{figure*}

\begin{figure*}[!htbp]
\centering
\includegraphics[width=0.44\textwidth]{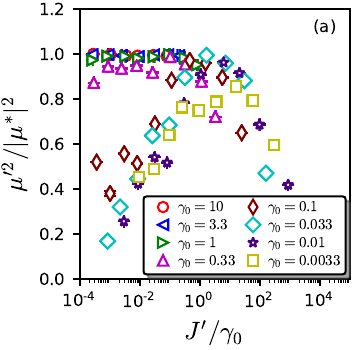}
\includegraphics[width=0.45\textwidth]{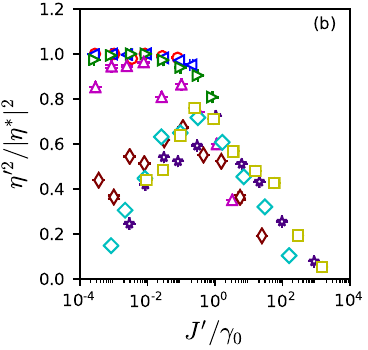}
  \caption{(a)~$\mu'^2/|\mu^{*}|^2$ and (b)~$\eta'^2/|\eta^{*}|^2$ as a function of $J'/\gamma_0$ for the suspensions composed of frictionless particles at various $\gamma_0$ as indicated by the legend.}
  \label{fgr:decom-nf}
\end{figure*}

One of the most common measurement in characterising viscoelastic materials are the storage modulus $G'$ and loss modulus $G''$, or equivalently the viscous (real) component of complex viscosity $\eta'$ and the elastic (imaginary) component $\eta''$. In this section, we present results of $\eta'$ and $\eta''$, as well as $\mu'$ and $\mu''$, previously defined in section \emph{Suspensions Rheology}. In Fig.~\ref{fgr:mu-ve} and \ref{fgr:mu-ve-nf}, we show how $\mu'$ and $\mu''$ vary with $J'$ for suspensions composed of frictional and frictionless particles respectively. The $\mu'(J')$-rheology are in both cases similar in behaviour to the steady shear curve. The main difference is that the curves shift to lower values compared to steady-state as $\gamma_0$ decreases. Fig.~\ref{fgr:mupc}(a) shows $\mu'_c$ as a function of $\gamma_0$ where we see a similar behaviour as for $|\mu*|_c$ with slightly lower values in the low $\gamma_0$-branch. \newline
The elastic component $\mu''$ is, on the other hand, the smallest, for a given $J'$ value, at large $\gamma_0$ values (see Fig.~\ref{fgr:mu-ve}(b)).
Several of the  $\mu''$-curves show a non-monotonic trend as a function of $J'$. The location of the minimum seems to vary with $\gamma_0$, leading to a non-monotonicity also in $\mu''_c$ as a function of $\gamma_0$ for a fixed $J'$, as seen in Fig.~\ref{fgr:mupc}(b) where we have estimated $\mu''_c$ to be equal to the value of $\mu''$ at our lowest $J'$ studied values. For the largest $\gamma_0$ the elastic component is close to zero, in agreement with what one would expect for a suspension with steady-shear rheology (\emph{i.e.}~an almost pure viscous response). 
% possibly showing a non-monotonicity around $\gamma_0=0.1$.
%shows a different behavior.
%The frictional and frictionless suspensions share similar behaviour in general. 
%The viscous component $\mu'$ behave essentially the same as for the steady-shear, the difference is that the curves shifts to lower values as the $\gamma_0$ decreases.
%At large $J$, $\mu''$ increases as $\sim J^{0.1}$ in both frictional and frictionless cases with more pronounced increase at smaller $\gamma_0$. At small $J$, $\mu''$ reaches plateaus in both frictional and frictionless cases. A non-monotonic dependence of the plateau values on $\gamma_0$ is observed, where the plateau value increases from a value close to zero at large $\gamma_0$, passes maximum at around $\gamma_0=0.0333$ and then decreases again at lower $\gamma_0$. At $\gamma_0=0.0033$, $\mu''_c\simeq 0.2$ for the frictional and $\mu''_c\simeq 0.05$ for the frictionless cases. 
%The elastic component $\mu''$ first decrease to a minimum and then further increase as $J$ increases at all $\gamma_0$. At large strain magnitude $\gamma_0=10$ and $3.3$, $\mu''$ decrease to roughly $0$ as $J$ decrease, while at smaller $\gamma_0$, $\mu''$ saturate at finite values. As for viscous component $\mu'$, one see that the values of $\mu'$ saturate at small $J$ at all $\gamma_0$ . The plateau values decreases with $\gamma_0$, which is the same trend as we found in $|\mu^{*}|$.\\
\newline
Fig.~\ref{fgr:decom} shows the relative importance of the viscous/elastic components to the overall complex quantities ($\mu'^{2}/|\mu^{*}|^{2}=1-\mu''^{2}/|\mu^{*}|^2$, $\eta'^2/|\eta^{*}|^2=1-\eta''^2/|\eta^{*}|^2$) as functions of $J'/\gamma_0$ for the suspensions composed of frictional particles. At large strain magnitudes ($\gamma_0\geq 1$), we have almost a pure viscous response. As $\gamma_0$ decreases, the elastic response becomes increasingly important. This is agreement with previous LAOS results of yielded dense Brownian suspensions in the high P\'eclet-regime[\onlinecite{Koumakis13}].
By normalising the viscous number $J'$ by $\gamma_0$, we obtain a decent collapse of the data for $\eta'^2/|\eta^{*}|^2$ and $\gamma_0\leq0.1$, as seen in Fig.~\ref{fgr:decom}(b). We do not observe the same collapse, however, for $\mu'^{2}/|\mu^{*}|^{2}$ (see Fig.~\ref{fgr:decom}(a)), instead we see that the peak values (if any) are $\gamma_0$-dependent.
%At $\gamma_0 > 0.33$, $\mu'/|\mu^{*}|$ and $\mu''/|\mu^{*}|$ are $1$ and $0$ respectively at large $J/\gamma_0$; for $\eta'/|\eta^{*}|$ one see a slight decrease at $\gamma_0=1$ and for $\eta''/|\eta^{*}|$ one see a slight increase at $\gamma_0=1$. At small $J/\gamma_0$, one see a clear split between different strain magnitude $\gamma_0$. At $\gamma_0\geq1$, the viscous components ($\mu'/|\mu^{*}|$ and $\eta'/|\eta^{*}|$) reach plateau which equal one (or equivalently the elastic components ($\mu''/|\mu^{*}|$ and $\eta''/|\eta^{*}|$) reach plateau with values close to zero). This suggests that at large strain magnitude, the response of the suspensions to oscillation is mostly viscous at low viscous number. At $\gamma_0\leq 0.1$, on the other hand, the response of the suspensions to oscillation is clearly viscoelastic. The viscous components first reach maximum and then decrease as $J/\gamma_0$ further decreases (the elastic component first decrease to a minimum and then increase). For $\eta'/|\eta^{*}|$ and $\eta''/|\eta^{*}|$ (Fig.~\ref{fgr:decom}(c) and (d)) the difference between different $\gamma_0$ is minor, \textit{i.e.}~they are within error bars with the peak values around $0.8$ for $\eta'/|\eta^{*}|$ and $0.6$ for $\eta''/|\eta^{*}|$ at all $\gamma_0\leq 0.1$. For $\mu'/|\mu^{*}|$ and $\mu''/|\mu^{*}|$ (Fig.~\ref{fgr:decom}(a) and (b)), the peak values seem to be $\gamma_0$ dependent. 
%At $\gamma_0=0.33$, one see a transition from clear viscoelastic response to more viscous-dominated response as $J/\gamma_0$ decreases. 
At small $J'/\gamma_0$ values, one sees a difference in the rheological behaviour depending on the value of $\gamma_0$. As $\gamma_0>0.1$ we see that the behaviour dominantly viscous as we approach its corresponding shear jamming point ($J'/\gamma_0 \to 0$). At $\gamma_0<0.33$, the opposite is true;  its elastic response dominates the shear-jamming. This transition correlates well with a change in the shear-jamming packing fraction for the frictional case. We do not attribute the emergence of elastic response to the elasticity of the particles as increasing the spring constants lead to the same curve (see the SI). Instead, we interpret this emergence of elasticity as coming from a not fully developed restructuring of the microstructure upon shear-reversal. A restructuring upon shear-reversal takes, in general, a finite and small strain, typically of the order of one, to be completed [\onlinecite{blanc2011local,Lin15,peters2016rheology}]. \\
%viscous component (or equivalently the elastic component) between different $\gamma_0$, which seems to be correlated to the shift in $\phi_c$.
%As seen in Fig.~\ref{fgr:phi}, at $\gamma_0\geq1$ the values of $\phi_c$ are the same as for the steady-shear case; at $\gamma_0\leq 0.1$ $\phi_c$ shift to higher values; at $\gamma_0=0.33$ $\phi$ deviate from the steady-shear case yet $\phi_c$ remain unchanged compared with the steady-shear case. 
Surprisingly, frictionless suspensions show a similar change in their rheological behaviour, see Fig.~\ref{fgr:decom-nf}, even though there is no clear shift in $\phi_c$ for these cases.
Hence, while there might be a correlation between the shift in $\phi_c$ and the transition from viscous to viscoelastic or even an elastic response as $\gamma_0$ decreases, there is no clear causality between these two observations. 
%Fig.~\ref{fgr:decom-nf} show a similar split between large and small $\gamma_0$ at small $J/\gamma_0$ for the frictionless suspensions even though there is no clear shift in $\phi_c$ for the frictionless suspensions. 
Comparing Fig.~\ref{fgr:decom-nf} to Fig.~\ref{fgr:decom}, we see a similar behavior of both the frictional and frictionless cases, with slightly worse collapsed of $\eta'^2/|\eta^{*}|^2$ as a function of $J'/\gamma_0$ for the low $\gamma_0$ values.
\subsubsection{Effect of lubrification forces and fluid stresses}
So far, we have neglected both contributions from lubrification forces and the fluid stress itself to the overall shear stress. The latter can be quite simply be corrected by adding a term $\eta_f$ to the loss modulus. This will mostly affect the curves where the complex viscosity (without the fluid stress) is small; see, \emph{e.g.,}~Fig.~\ref{fgr:eta} above and Fig.~1 in the SI. 
As a first approximation, one could argue that the lubrification forces acts as a mean-field viscosity with increased \emph{effective} fluid viscosity, $\eta_f^{\rm eff}$, as a result. Assuming that this \emph{effective} viscosity has a moderate dependence on the packing fraction, one could rationalise this effect by a redefinition of $J'$ as well as a simple upscaling of the stresses $\sigma \propto \eta_f^{\rm eff}$ rather than $\sigma \propto \eta_f$. However, as the packing fraction is increased, the stresses become dominated by the contact stresses, and our approximation of neglecting stresses from lubrification forces and the fluid itself becomes increasingly accurate. That the viscoelastic response is predominantly caused by the contact stresses can be, for example, seen in the work by Ness, Xing, and Eiser [\onlinecite{Ness17}] (see their Fig.~3). For completeness, we provide additional data in the SI, where lubrification forces have been accounted for. As stated above, this mostly affects the low packing fraction region (equivalent to the high $J'$-regime) and does not change any of our conclusions regarding what happens when approaching the shear-jamming regime. 
%At large $J/\gamma_0$, one see a similar collapsing of data at different small $\gamma_0$ for the frictionless suspensions with the same scaling as for the frictional suspensions, as shown by the black dashed lines in Fig.~\ref{fgr:decom-nf}. The main difference between the frictional and frictionless suspensions is that one does not see a $\gamma_0$-dependence on the peak values of $\mu'/|\mu^{*}|$ and $\mu''/|\mu^{*}|$ for the frictionless suspensions.

\subsection{MICROSCOPIC STRUCTURE}
\subsubsection{Number of contacts}
\begin{figure*}[!htbp]
\centering
\includegraphics[width=0.33\textwidth]{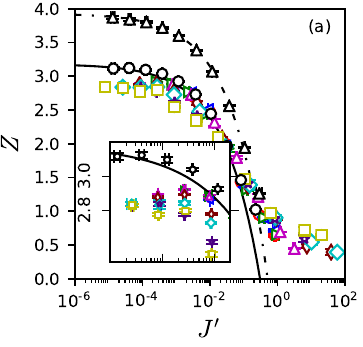}
\includegraphics[width=0.33\textwidth]{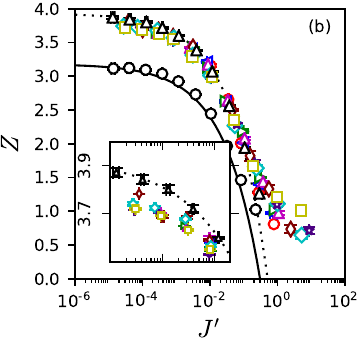}
\includegraphics[width=0.315\textwidth]{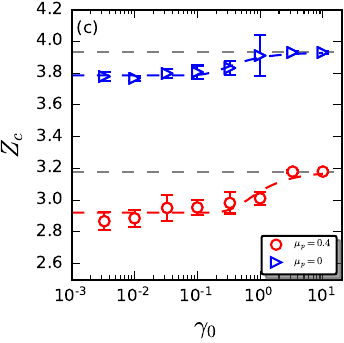}
  \caption{The number of contacts per particle $Z$ as a function of the viscous number $J'$ at various $\gamma_0$ for (a)~frictional and (b)~frictionless suspensions. The colours and symbols are the same as in Fig.~\ref{fgr:eta}. The black symbols correspond to the steady-shear conditions (circles for the frictional suspensions and triangles for the frictionless suspensions), and the black lines are plots of the constitutive laws for the steady-shear cases. The constitutive laws are given as $Z=Z_c-a_ZJ^{n_Z}$, with $a_Z=5.2$, $n_Z=0.41$ for both the frictional and frictionless cases, $Z_c=3.18$ for the frictional case and $3.93$ for the frictionless case. The insets are the zoomed-in figures of the main figures close to $J'\to 0$. In (c) $Z_c$ as a function of $\gamma_0$ for both the frictional and frictionless suspensions; the coloured dashed lines are best fits of the phenomenological function $Z_c=Z_c^{SS}[1-c_1\tanh(c_2/\gamma_0)]$, where $Z_c^{SS}$ is the values of suspensions under steady shear with $Z_c^{\mathrm{SS}}=3.18$ for the frictional and $\mu_c^{\mathrm{SS}}=3.93$ for the frictionless cases, as indicated by the grey dashed horizontal lines. $c_1$ and $c_2$ are two fitted parameters, which values are given in the text.}
  \label{fgr:z}
\end{figure*}

Fig.~\ref{fgr:z} shows how the number of contacts per particle $Z$ evolves as $J'$ varies for a suspension composed of (a)~frictional or (b)~frictionless particles at different $\gamma_0$. In general, we find that $Z$ follows the same curves as for steady-state. The exception is again at small $\gamma_0(\leq 1)$, where we get smaller $Z$ values compared to steady-state's $Z_c^{\mathrm{SS}}$. This exception can be better be seen in the insets of the figure. As before, we estimate the values of $Z_c$ at different $\gamma_0$ by fitting the data to $Z=Z_c-a_Z(J')^{n_Z}$ with $J'<10^{-2}$, where $Z_c$, $a_Z$ and $n_Z$ are three free fitting parameters. The results are presented in Fig.~\ref{fgr:z}(c) as functions of strain magnitude $\gamma_0$, where we find a similar behaviour as for the complex macroscopic friction coefficients (see Fig.~\ref{fgr:mu}(c)). We again fit the data to a phenomenological hyperbolic function $Z_c=Z_c^{\rm{SS}}[1-c_1\tanh(c_2/\gamma_0)]$, where $Z_c^{\rm{SS}}$ is the value of $Z_c$ for steady-shear and equal to $3.18$ for the frictional particles and $3.93$ for the frictionless case; $c_2$ shows at around which $\gamma_0$-value one sees a shift in the number of contacts per particle (at shear-jamming). For the frictional case we find $c_2=0.53\pm0.21$ and for the frictionless case $c_2=0.23\pm0.05$. Interestingly these shifts occur at considerably larger $\gamma_0$ values than we have seen previously for $[\mu^{*}|$; hence there is no direct link between these two quantities. Instead, the change is better correlated with the change in $\phi_c$ (if any). $Z_c^{\rm{SS}}(1-c_1)$ gives the $Z_c$ value as $\gamma_0\to 0$, where we find $c_1=0.08\pm0.007$ (frictional) and $c_1=0.04\pm0.002$ (frictionless) leading to $Z_c=2.96$ (frictional) and $Z_c=3.77$ (frictionless) for small $\gamma_0$-values. These values show that the frictional case is isostatic (\emph{i.e.,}~very close to 3) while the frictionless is hypostatic (\emph{i.e.,}~considerable less than 4). Rattlers have, however, not been excluded in these estimates. Excluding rattlers would increase the number of contacts (per no rattling particle) slightly. \newline
%\begin{table}[h]
%\begin{center}
%\caption{The values of $c_1$ and $c_2$ for the frictional and frictionless cases}
%\begin{tabular}{ |c|c|c| } 
 %\hline
 % & frictional & frictionless \\ \hline 
% $c_1$ & $0.07\pm0.007$ & $0.53\pm0.21$ \\ 
% $c_2$ & $0.04\pm0.002$ & $0.23\pm0.05$ \\ 
% \hline
%\end{tabular}
%\end{center}
%\end{table}

\begin{figure*}[!htbp]
\centering
\includegraphics[width=0.33\textwidth]{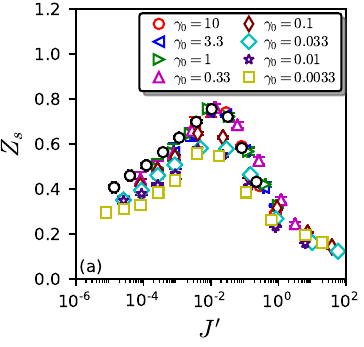}
\includegraphics[width=0.33\textwidth]{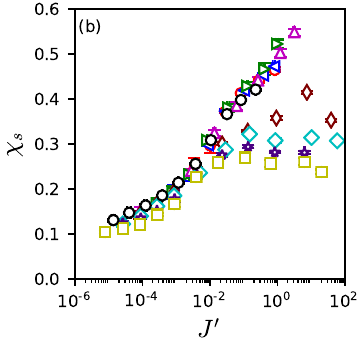}
  \caption{(a)~Number of sliding contacts per particle $Z_s$ and (b)~fraction of sliding contacts $\chi_s$ as a function of the viscous number $J'$ at various $\gamma_0$ as indicated in the legends for the suspensions composed of frictional particles. The black symbols are data for the steady-shear.}
  \label{fgr:zs}
\end{figure*}

Frictional particles at contact can either undergo sliding or rolling motions. For $\mu_p=0.4$, the frictional sliding dominates the energy dissipation close to shear jamming[\onlinecite{trulsson2017effect}]. In addition, the frictional sliding contributes to a higher shear jamming packing fraction than for suspensions composed of frictional particles without sliding[\onlinecite{van2009jamming,singh2020shear}]. This could explain why we see an increased packing fraction in the frictional case but not the frictionless (where all contacts can be regarded as sliding), as oscillations could alter the fraction of sliding particles.
Fig.~\ref{fgr:zs}(a) shows the number of sliding contacts per particle $Z_s$ as a function of $J'$ at various $\gamma_0$. At $\gamma_0 \geq 0.33$, the curves are again overlapping with the steady-shear one. As $\gamma_0$ is lowered, one can see a decrease in $Z_s$ and the fraction of sliding contacts $\chi_s=Z_s/Z$; see Fig.~\ref{fgr:zs}(b). A lower fraction of sliding contacts would indicate a smaller packing fraction in general, in contradiction to what is found (see Fig.~\ref{fgr:phi}). Hence, the increased packing fractions are not explained by a decrease in the number of sliding contacts per particle or fraction of them but rather by reducing the overall number of contacts.  
%Neither does it offers an explaination to why we have shift in $\phi_c$ for low $\gamma_0$ values in the frictional case but not the frictionless.
%This is a possible explanation for the shift of $\phi_c$ we observe in Fig.~\ref{fgr:phi}. The oscillatory shear decreases the sliding contacts in the suspensions which allows the particles to further compact, leading to a higher shear jamming packing fractions $\phi_c$.
 %For the suspensions composed of frictionless suspensions, there is no sliding contact at all so that the particles are not able to further compact by having less sliding contacts. Therefore we do not see any clear increase in $\phi_c$ in the frictionless cases.  Fig.~\ref{fgr:zs}(b) shows . The same trend is found, with smaller $\gamma_0$ leading to lower $\chi_s$. 
\subsubsection{Geometrical contact fabric}
\begin{figure*}[!htbp]
\centering
\includegraphics[width=0.32\textwidth]{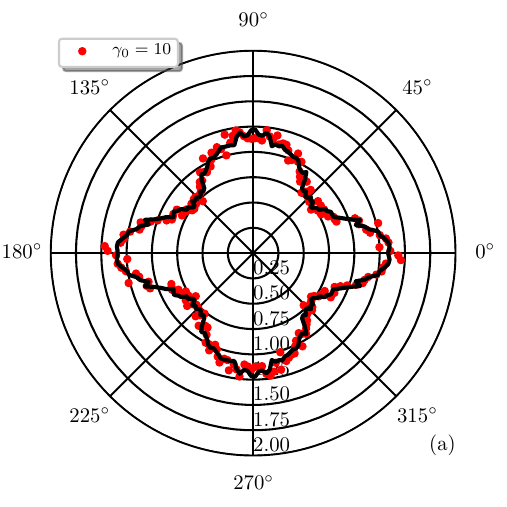}
\includegraphics[width=0.32\textwidth]{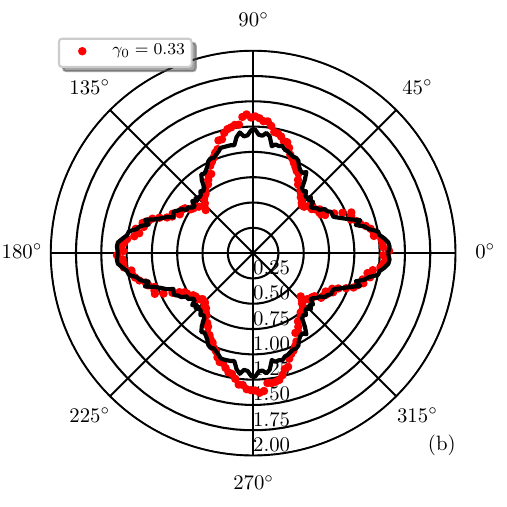}
\includegraphics[width=0.32\textwidth]{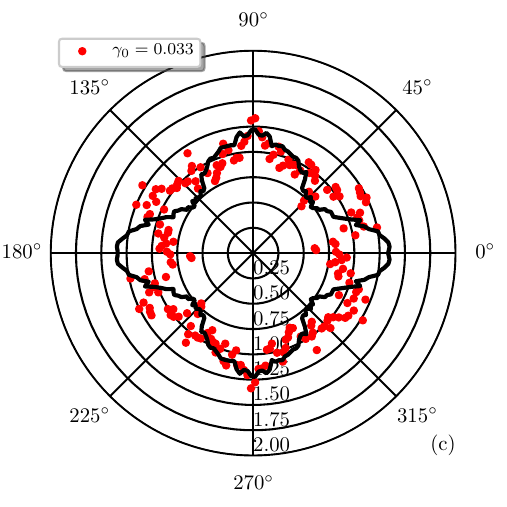}
  \caption{The geometrical contact fabric of the suspensions composed of frictional particles at $J'\simeq 3\cdot 10^{-3}$ and (a)~$\gamma_0=10$, (b)~$\gamma_0=0.33$ and (c)~$\gamma_0=0.033$. The black lines give the steady-shear results at the same $J'$. The corresponding $\phi$ are (a)~$0.770$, (b)~$0.797$, and (c)~$0.832$.}
  \label{fgr:fabric}
\end{figure*}
\begin{figure*}[!htbp]
\centering
\includegraphics[width=0.32\textwidth]{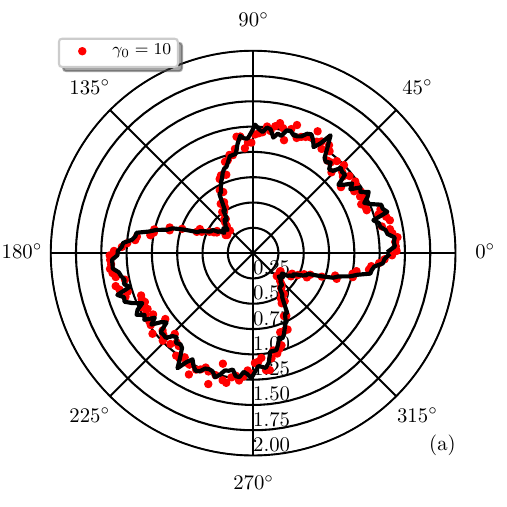}
\includegraphics[width=0.32\textwidth]{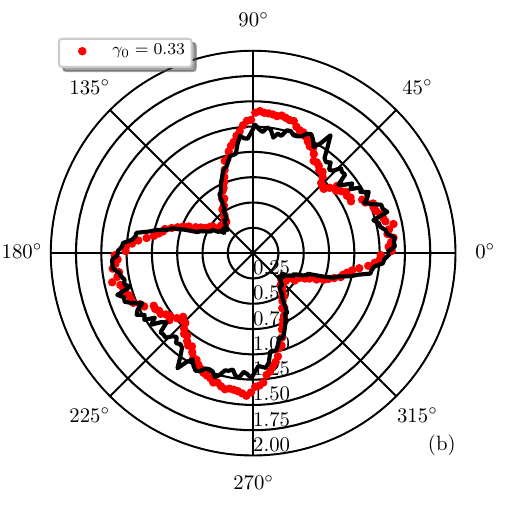}
\includegraphics[width=0.32\textwidth]{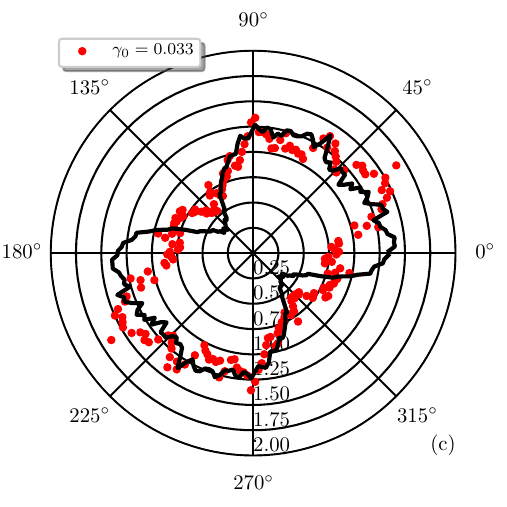}
  \caption{The same as in Fig.~\ref{fgr:fabric} but along the shear direction.} %(\textit{i.e}~without symmetrisation).}
  \label{fgr:fabric_shear}
\end{figure*}
The geometrical contact fabric gives the probability of having a contact at a certain angle, where the angle is between the vector, which connects the particles at contact and the axis parallel to the walls[\onlinecite{radjai1998bimodal}].
The shear-rate averaged geometrical contact fabric is calculated, in a similar way as for the other shear-rate averaged quantities, as
\begin{equation}
\xi(\theta) = \frac{\int_0^{2\pi/\omega} \xi(\theta,t)|\dot\gamma(t)|dt}{\int_0^{2\pi/\omega} |\dot\gamma(t)|dt}.
\end{equation}
We also consider an alternative contact fabric which accounts for that the shear direction reverses. This geometrical fabric measures the fabric relative to the shear direction), and which we denote as $\xi'(\theta)$, is defined as
\begin{equation}\label{eq:xi}
\xi'(\theta)=
\begin{cases}
\xi(\theta) ,& \dot\gamma > 0;\\ \xi(\pi-\theta),& \dot\gamma \leq 0;
\end{cases}
\end{equation}
In Fig.~\ref{fgr:fabric} and \ref{fgr:fabric_shear}, we show how the shear-rate averaged geometrical contact fabrics  $\xi(\theta)$ is affected by the oscillatory shear at (a)~$\gamma_0=10$, $\gamma_0=0.33$, and $\gamma_0=0.033$ for the suspensions composed of frictional particles at $J'\simeq 3\cdot10^{-3}$,  without and with accounting for the shear-reversal respectively.
The black lines indicate how geometrical contact fabrics would look if the suspensions behave the same as under steady-shear. The black lines in Fig.~\ref{fgr:fabric} are symmetrised by reflection, accounting for the shear-reversal in our system.
At $\gamma_0=10$, the geometrical fabrics are almost identical to the steady-state both when accounting for shear-reversal direction and not. The geometrical fabrics at $\gamma_0=0.33$ show only a minor difference than those at $\gamma_0=10$. As $\gamma_0$ is further lowered (at $\gamma_0=0.033$), the contact fabrics change from four-fold symmetry to possibly a six-fold in Fig~\ref{fgr:fabric}. The geometrical fabric along the shear direction and at $\gamma_0=0.033$ is altered but remains anisotropic with high probabilities along the compression axis. The microstructure is mostly altered along the extension axis (increased probability) and the shear direction (decreased probability).
Similar trends can be found for the normal and tangential fabrics and higher $J'$ values (see the SI). In general, there is no big difference between the frictional and the frictionless cases, neither in shape nor in values, as shown in the SI.

\section*{Conclusions}

In this work, we have studied the behaviour of dense suspensions composed of either frictional or frictionless particles in oscillatory shear flows at imposed pressures. We have shown that the $\mu(J)$-rheology for non-Brownian suspensions can be generalised to oscillatory shear flows, allowing us to scan the viscoelastic properties of these suspensions in the framework a $\mu(J)$-rheology. We found that at large strain magnitude $\gamma_0\geq 1$, the rheological behaviour can be well-described by their corresponding rheologies in steady-shear. We interpret this as the suspensions in those cases have sufficient time to recover the steady-state microstructure at each shear-reversal. At smaller strain magnitudes, the rheology deviates from the steady-shear case (\emph{i.e.,}~does not respect the Cox-Merz rule). 
In both the frictional and frictionless cases, we observe a transition from a pure viscous to a viscoelastic response as strain magnitude $\gamma_0$ decreases. 
For both frictional and frictionless cases, we find that both the complex macroscopic friction coefficient $|\mu^{*}|$ and the number of contacts $Z$ decrease compared to steady-shear for $\gamma_0<0.3$ and close to their corresponding shear jamming points. Although there is no clear indication of the formation of reversible states at small $\gamma_0$ [\onlinecite{Forthcoming}], a consequence of our pressure imposed set-up, the lowered number of contacts and friction coefficient, as well as the increased packing fraction (frictional case), indicate that the particles partially ``self-organise'' or ``re-organise'' themselves[\onlinecite{corte2008random,pine2005chaos}], even though there is no clear signature in the contact fabric as $\gamma_0$ is lowered. 
As previously reported[\onlinecite{Dong2020-1}], the shear jamming packing fractions are found to shift to higher values at small strain magnitudes in the frictional cases, $\phi_c^{\mathrm{f,SS}} < \phi_c < \phi_c^{\mathrm{nf,SS}}$. This despite the fact that the fraction of sliding contacts is (mildly) lowered. 
For the frictionless cases, we see no significant increase in $\phi_c$, although the critical number of contacts is lowered with decreasing strain amplitude.
While lowering the strain amplitude yields a lower critical stress ratio and higher packing fractions (frictional case), this also leads to an emergence of elastic (or viscoelastic) behaviour.
For strain amplitudes $\gamma_0<0.33$ we find that the shear-jamming points are elastic, while for $\gamma_0\geq1$ they are viscous. 

\section*{Conflicts of interest}
There are no conflicts to declare.

\section*{Acknowledgements}
MT acknowledges financial support by the Crafoord Foundation (20190650). The simulations were performed on resources provided by the Swedish National Infrastructure for Computing (SNIC) at the centre for scientific and technical computing at Lund University (LUNARC). 

\section*{Author Contributions}
MT conceptualised the project. JD carried out the numerical simulations. JD and MT analysed the data and wrote the paper.

%\balance


%%%REFERENCES%%%
%\nocite{*}
\bibliography{ref} 

\end{document}